\documentclass[pra,twocolumn]{revtex4}

\usepackage[dvips]{graphicx}%
\usepackage{amsmath}
\usepackage{bm}
\begin{document}
%%%%%%%%
  \small
%%%%%%%%

\title{Verification of quantum-domain process using two non-orthogonal states}

\author{Ryo Namiki}% / \today
\affiliation{Department of Physics, Graduate School of Science, Kyoto University, Kyoto 606-8502, Japan}

%\affiliation{CREST Research Team for Photonic Quantum Information, Division of Materials Physics, Department of Materials Engineering Science, Graduate school of Engineering Science, Osaka University, Toyonaka, Osaka 560-8531, Japan}
\date{\today}%March 17, 2007}%
\begin{abstract}

If a quantum channel or process cannot be described by any measure-and-prepare scheme, we may say the channel is in \textit{quantum domain} (QD) since it can transmit quantum correlations. The concept of QD clarifies the role of quantum channel in quantum information theory based on the local-operation-and-classical-communication (LOCC) paradigm: The quantum channel is only useful if it cannot be simulated by LOCC.  
We construct a simple scheme to verify that a given physical process or channel is in QD by using two non-orthogonal states. We also consider the application for the experiments such as the transmission or storage of quantum optical coherent states, single-photon polarization states, and squeezed vacuum states.

\end{abstract}

%\pacs{03.67.Dd, 42.50.Lc} 
%\keywords{quantum cryptogrphy}
\maketitle

%\newpage
%%%%%%%%%%%%%%%%%%%%%   part 1 %%%%%%%%%%%%%%%%%%%%%%%%%%%%%%%%%%%%%%%%%%%%%%%%
%%%%%%%%%%%%%%%%%%

%\small 

%Abstract: \textit{}
 \section{Introduction}
A transfer of an unknown state is a primary object of quantum information science. Since the phrase ``unknown state'' suggests that the physical system %
 is possibly entangled with another system, the foundation of this object can be related with the change of quantum correlation thorough the transfer process.
Associated with the maintenance of the inseparability, a distinguishing class of local operations is the so-called \textit{entanglement breaking} (EB) channel that breaks any entanglement, i.e., a local operation $\Phi$ is EB if  $\openone_A\otimes  \Phi_B ( \hat \rho_{AB})$ is a separable state for any state $\hat \rho_{AB} $ \cite{16,17}. It is well-known that an operation is EB if and only if it can be written as a \textit{measure-and-prepare} (M\&P) scheme that assigns the output states based on the classical data obtained by the measurement of the input states.
When a process is not a M\&P scheme, there exists an entangled state that maintains inseparability after the process and it can transmit non-classical correlations.
Hence, it is natural to say that the process is in \textit{quantum domain} (QD) if  the process is not a M\&P scheme. This poses clear distinction between quantum processes and classical processes firmly based on the maintenance of quantum correlation.  
In quantum information theory, the local operation and classical communication (LOCC) is set free to use, hence, the quantum channel is only useful if it cannot be simulated by LOCC. 
The assurance of QD processes tells us that a given quantum channel is different from any LOCC channel. 
 Subsequently, the criterion for QD processes has been quantum benchmark of the experimental success of core physical processes, such as transmission or storage of quantum states % 
 \cite{Ham05,Bra00,namiki07}. %,
 Mathematically, the set of QD channels is connected with a set of inseparable states by Jamiolkowski isomorphism \cite{16,17,Hua06}, and the concept of QD is considered to be the inseparability of quantum channels \cite{qdc}.

In principle, one can determine a given process by the process tomography, and check the necessary and sufficient condition for EB channel \cite{16,17}. However, tomographic reconstruction is not always easy to perform. Assuming a practical channel and a limited set of experimental parameters, several QD criteria have been proposed %
 associated with the quantum key distribution (QKD) \cite{Gro03b,Rig06,Has08}. Thereby, the problem is rather identified as a type of entanglement verification/detection and the formulations are deeply related with the entanglement witness \cite{Cur04}. %  
On the other hand, it might be more direct to demonstrate better-than-classical performance by introducing certain figure of merit when one shows the success of experiments. %
A familiar approach is to investigate the average fidelity of the process with respect to an ensemble of states \cite{Pop94,Mass95,Fuc03}. If one can find the upperbound of the average fidelity achieved by the M\&P schemes, surpassing the bound is a sufficient condition of QD processes \cite{Bra00,Fuc03,Ham05,namiki07}. The optimization problem of the average fidelity is also investigated in the state estimation and optimal cloning \cite{Bae06,rmp-clone}. %

 Aside from the quantum inseparability, an assurance of genuine quantum devices could be that not only a set of  orthogonal states but also a set of their supperpositions is coherently transferred. As in the sprit of the two-state QKD scheme \cite{B92}, the coherence can be demonstrated by testing with two non-orthogonal states, and it would be important to construct an experimentally simpler verification scheme of QD processes as well as  a solid foundation on the primary object. %
Based on the transmission of binary coherent states and quadrature measurements, a verification scheme is developed in \cite{Rig06,Has08}. A general approach that concerns the average fidelity for two non-orthogonal states is found in Ref. \cite{Fuc03}.

In this paper, we construct a simple verification scheme of QD processes using two non-orthogonal states as a variant of  \cite{Fuc03}. 
The setup %of the verification scheme
 is as follows:
A pair of pure states $| \psi_\pm \rangle $ with non-zero overlap is prepared and experiences a physical process $E$. Suppose that the process $E$ converts the input states as $ \hat \rho_\pm = E(| \psi_ \pm   \rangle \langle \psi_ \pm  | )$, %$| \psi_ \pm   \rangle \to \hat \rho_\pm $,
 and the projection probabilities of the output onto the pair of target states $| \psi _\pm ' \rangle $,  %to be %, respectively.  and that we obtain the values of overlap 
say $b=  \langle \psi_+' |\hat \rho_+ |\psi_+' \rangle  $ and $ a= \langle \psi_-' |\hat \rho_- |\psi_-' \rangle  $,  are measured. We show the condition on $a$ and $b$ %the two values 
that ensures that the process is in QD. %The measurements of $a$ and $b$ are directly connected with the photon detection in contrast with the quadrature measurement employed in \cite{Rig06,Has08}.
We derive the criterion in Sec. II and consider applications for quantum-optical experiments in Sec. III. We make a conclusion in Sec. IV. %

\section{Criterion for Quantum domain processes with two input states}
\subsection{Average fidelity and its classical boundary for transformation task of a set of states}

Any physical process is described by a completely positive trace-preserving (CPTP) map \cite{NC00}. We define the average fidelity on a process $E$ with respect to the transformation task from a set of input states $\{ | \psi_i \rangle \}$ to a set of \textit{target} states $\{| \psi_i' \rangle \}$ with a prior distribution $\{p_i\}$ \cite{namiki07} by 
\begin{eqnarray}
\bar F &=& \sum _i p_i \langle \psi_i'| E \left( |\psi_i \rangle\langle \psi_i |  \right) | \psi_i'\rangle.
\end{eqnarray} %where $p_i$ is the prior distribution of the input states.
% Different from the ordinary definition of the average fidelity, i.e.,  $| \psi_i \rangle \ =| \psi_i' \rangle \ $ for any $i$, it is possible that $\bar F$ does not reach unit even if $E$ is optimized over the CPTP map \cite{namiki07}. % 
The process is simulated by the M\&P schemes when we can write \begin{eqnarray}
 E \left( |\psi_i \rangle\langle \psi_i |  \right)  &=& \sum _k  \langle \psi_i | \hat M _k| \psi_i  \rangle
\hat \rho_k \end{eqnarray} where $\{ \hat M_k \}$ is a positive-operator valued measure (POVM) and $\hat \rho_k $ is a density operator.  
The classical boundary of the average fidelity for the task $\{ | \psi_i  \rangle \} \to \{| \psi_i' \rangle \} $ is defined by the optimization over the M\&P schemes: 
%\begin{eqnarray}F_{c} %\nonumber \\ &\equiv & \sup_{\{\hat M_k  \},\{\hat \rho_k \} }\sum_{x,k} p_i \textrm{Tr}  \left( \hat M_k |\psi_i \rangle \langle \psi_i  |  \right)    \langle\psi_i'  |\hat \rho_k |\psi_i' \rangle   \label{classf} . \end{eqnarray}
%Task: $\{| \psi_i \rangle  \to | \psi_i' \rangle  \} $
%Ensemble of inputs: $\{p_i, | \psi_i \rangle   \} $
%POVM: $\{ \hat E_k \} $ 
%Assignment$\{k \to \hat \rho_k  \} $
\begin{eqnarray}
F_c &\equiv& \sup_{\hat M_k, \hat \rho_k} \sum_i \sum_k  p_i \langle \psi_i |\hat M_k |\psi_i \rangle  \langle \psi_i' |\hat \rho_k |\psi_i' \rangle  \nonumber \\
&=&\sup_{\hat M_k}  \sum_k  \left \|  \sum_i  p_i \langle \psi_i |\hat  M _k |\psi_i \rangle   |\psi_i' \rangle
 \langle \psi_i' |   \right\|_\infty  \nonumber\\ 
 &\equiv&  \sup_{\hat M_k }  \sum_k \left \|  \hat A_k  \right\|_\infty .  \end{eqnarray} 
where $\| \cdot \|_\infty $ denotes operator norm.
 We can verify the process $E$ is in QD if measured $\bar F$ exceeds $F_c$. Note that the optimization problem reduces to the problem of the minimum error discrimination (MED) \cite{Moc06} when  $ \langle\psi_i '|\psi_j ' \rangle = \delta _{i,j}$. In this case we can see that $F_c \ge \bar F$ for any CPTP, and the orthogonal-target task is not useful to make the QD verification scheme. %,
 An interesting point is that %we can find a role of 
 the quantum correlation gains the score if the problem moves from the point of the MED problem. With the non-orthogonality between the target states $| \langle\psi_i '|\psi_j ' \rangle  |$ as a parameter we can work on a unified framework that includes the two widely investigated class of the problems: the state estimation  $| \langle\psi_i' |\psi_j '\rangle  | = | \langle\psi_i |\psi_j  \rangle  |$ \cite{Fuc03,Mass95,Pop94,Bae06,rmp-clone} and the MED problem. The relation between the two problems was discussed in a different aspect \cite{Fuc03,Bar01}. %

\subsection{Classical boundary fidelity for two-state case}
We start the two-state case by denoting $i = \pm $ and all the relations between the states are described by two parameters % assume the relation of the states: 
\begin{eqnarray}
\gamma &\equiv &  |\langle \psi_+ | \psi_- \rangle |  \nonumber\\
\gamma' &\equiv&  |\langle \psi_+' | \psi_-' \rangle |    \label{gam} %( 0 \le \theta, \theta' \le \pi ) 
. 
\end{eqnarray}
%\begin{eqnarray}\langle \psi_+ | \psi_- \rangle &=& \cos \theta \equiv  \gamma   \\\langle \psi_+' | \psi_-' \rangle &=& \cos \theta ' \equiv  \gamma',  ( 0 \le \theta, \theta' \le \pi ) . \end{eqnarray}
% 
The upperbound $F_c$ can be obtained by following the discussion given by Fuchs and Sasaki \cite{Fuc03} where $\gamma = \gamma' $, however, the proof of the bound is somewhat complicated. 
Here, we provide a different derivation of $F_c$ and the proof is quite simpler. %proof that include $\gamma = \gamma' $ case.

By choosing the orthogonal basis of the target states $  | \pm ' \rangle \equiv (| \psi_+  ' \rangle \pm |\psi ' _- \rangle )/  \sqrt{2 (1\pm \gamma')}$, we can write   
\begin{eqnarray}
\hat A_k    = \frac{1}{2}\left(
  \begin{array}{cc }
 \rho_k (1+ \gamma') & \Delta _k \sqrt{1- \gamma'^2 }    \\
\Delta _k \sqrt{1- \gamma'^2 }  & \rho_k (1- \gamma')     \\  \end{array}
\right).
\end{eqnarray} 
where  we defined
\begin{eqnarray}
 \hat \rho  & \equiv& p_+  |\psi_+  \rangle\langle\psi_+ | +   p_-  |\psi_- \rangle\langle\psi_- |     \\
\hat  \Delta   & \equiv&  p_+  |\psi_+  \rangle\langle\psi_+ | -   p_-  |\psi_- \rangle\langle\psi_- |         \end{eqnarray} 
and 
\begin{eqnarray}
 \rho_k   & \equiv&  \textrm{Tr} \hat M_k  \hat \rho  \\
 \Delta _k  & \equiv&  \textrm{Tr}\hat M_k  \hat \Delta . \end{eqnarray} 
Then we have \begin{eqnarray} \left \|  \hat A_k  \right\|_\infty = \frac{1}{2}\left(
  \rho_k  + \sqrt{ {\rho_k} ^2   \gamma'^2 + {\Delta _k }^2 (1- \gamma'^2 ) }  \right)  \end{eqnarray} and 
\begin{eqnarray}
F_c % &= & \sup_k \sum_k \frac{1}{2}\left(  \rho_k  + \sqrt{ {\rho_k} ^2   \gamma'^2 + {\Delta _k }^2 (1- \gamma'^2 ) }  \right) \nonumber\\
%  &= & \frac{1}{2}\left(  1+ \sup_k \sum_k  \sqrt{ {\rho_k} ^2   \gamma'^2 + {\Delta _k }^2 (1- \gamma'^2 ) }  \right) \nonumber \\
  &= & \frac{1}{2}\left(
  1+ \sup_{\hat M_k}  \sum_k  \rho_k    \sqrt{   \gamma'^2 + (1- \gamma'^2 )  {\Delta _k }^2/ {\rho_k} ^2  }  \right) \nonumber . \\ \label{fc1}
%&\le & \frac{1}{2}\left(  1+ \sup_{\hat M_k}   \sqrt{ \gamma'^2  +  (1- \gamma'^2 )  \sum_k {\Delta _k }^2/ \rho_k  }   \right)  \nonumber %\\%
%&\le & \frac{1}{2}\left( 1+   \sqrt{ \gamma'^2  +  (1- \gamma'^2 ) \left( P^2    \gamma ^2 +  1- \gamma ^2   \right)  }   \right)\nonumber \\%&\le & \frac{1}{2}\left(  1+   \sqrt{ \gamma'^2  +  (1- \gamma'^2 ) \left( P^2  \cos^2 \theta + \sin ^2 \theta \right)  }   \right)\nonumber \\ 
  %&= & \frac{1}{2}\left(  1+ \sup_k  \sqrt{\sum_k  \left\{ {\rho_k} ^2   \gamma'^2 + {\Delta _k }^2 (1- \gamma'^2 ) \right\} }  \right)\\
 % &= & \frac{1}{2}\left(  1+ \sup_k  \sqrt{  \gamma'^2 + (1- \gamma'^2 ) \sum_k {\Delta _k }^2  }  \right)\\
  \end{eqnarray}

%\begin{eqnarray}F_c &=&\sup_{\hat M_k}  \sum_k  \left \|  \sum_{i= \pm}  p_i \langle \psi_i |\hat  M _k |\psi_i \rangle   |\psi_i' \rangle \langle \psi_i' |   \right\|_\infty  \nonumber\\  &=&\frac{1}{2}\left(  1+ \sup_{\hat M_k}  \sum_k  \rho_k    \sqrt{   \gamma'^2 + (1- \gamma'^2 )  {\Delta _k }^2/ {\rho_k} ^2  }  \right) \nonumber   .  \end{eqnarray}

Let us choose the orthogonal basis of the input states $  | \pm \rangle \equiv (| \psi_+  \rangle \pm |\psi _- \rangle )/  \sqrt{2 (1\pm \gamma )}$, and define the Pauli operators by $\hat \sigma_0 =|+ \rangle \langle +| + |- \rangle \langle -|, \hat \sigma_z =|+ \rangle \langle +| - |- \rangle \langle -| , \hat \sigma_x =|+ \rangle \langle -| + |- \rangle \langle + | , \hat \sigma_y = i|- \rangle \langle + | -i |+ \rangle \langle -  |    $.
 Then, we can write 
\begin{eqnarray} 
\hat \rho &=&\frac{1}{2} \left( \hat \sigma_0 +  
 PG  \hat \sigma_x  \sin \phi_0+ \label{form1} 
  \frac{G}{P} \hat \sigma_z \cos \phi_0 
 \right) \\
\hat \Delta &=&\frac{1}{2} \left( P \hat \sigma_0 + G  \hat \sigma_x
  \sin \phi_0  + G \hat \sigma_z  \cos \phi_0  
  \right) ,
\end{eqnarray} 
where we defined
\begin{eqnarray}
%\vec  {\sigma }  &\equiv& (\hat \sigma_x, \hat \sigma_y,\hat \sigma_z)^t\\
P &\equiv&   p_+-p_-  \\ 
G&\equiv& \sqrt{P^2 \gamma ^2 +(1-\gamma^2)}\\
%K&\equiv& \sqrt{  {\gamma'} ^2 +(1-{\gamma'}^2)G^2}\\
\cos\phi_0&=& \frac{P \gamma}{G} \\
\sin\phi_0&=& \frac{\sqrt{ 1-\gamma 2} }{G} .
\end{eqnarray} 
%In what follows, we assume $P \ge 0$.
%\begin{eqnarray} \hat \rho &=&\frac{1}{2} \left( \hat \sigma_0 + \left(\begin{array}{c } PG \sin \phi_0  \label{form1}\\0 \\  G/P \cos \phi_0 \\ \end{array}\right) \cdot \vec  {\sigma }\right) \\\hat \Delta &=&\frac{1}{2} \left( P \hat \sigma_0 + G\left(  \begin{array}{c }  \sin \phi_0  \\0 \\  \cos \phi_0 \\ \end{array}\right) \cdot \vec  {\sigma }\right) ,\end{eqnarray} 
%where we defined\begin{eqnarray}\vec  {\sigma }  &\equiv& (\hat \sigma_x, \hat \sigma_y,\hat \sigma_z)^t\\P &\equiv&   p_+-p_-  \\ G&\equiv& \sqrt{P^2 \gamma ^2 +(1-\gamma^2)}\\\cos\phi_0&=& \frac{P \gamma}{G} \\\sin\phi_0&=& \frac{\sqrt{ 1-\gamma 2} }{G} .\end{eqnarray} 

Since  %both the coefficients of $\hat \sigma_y$ for 
$\textrm{Tr}( \hat \rho \hat \sigma_y )= \textrm{Tr} (\hat \Delta \hat \sigma_y )=0 $, we can choose the optimal POVM so that $\textrm{Tr}  (\hat M_k \hat \sigma_y) =0$ without loss of generality. Then, %Noting that POVM is alwayes decomposed to rank-1 projections, 
we can describe rank-1 POVM element $\hat M_k$ as a real vector in the Bloch sphere with a single parameter $\phi_k$, 
\begin{eqnarray}
\frac{\textrm{Tr}  (\hat M_k \hat \sigma_z)}{\textrm{Tr}   \hat M_k} &=& \cos (\phi _k +\phi _0  ) \\
\frac{\textrm{Tr}  (\hat M_k \hat \sigma_x) }{{\textrm{Tr}   \hat M_k}}&=& \sin (\phi _k +\phi _0  ). \label{to1}
\end{eqnarray} 
The condition of the POVM, $ \sum_k \hat M_k= \hat \sigma_0$, implies \begin{eqnarray}
\sum_k  \textrm{Tr}(\hat M_k \hat \sigma_z ) &=& \sum_k \textrm{Tr}   \hat M_k \cos (\phi _k +\phi _0  ) = 0\label{Pz}\\ \sum_k \textrm{Tr}(\hat M_k \hat \sigma_x ) \label{Px} &=&\sum_k \textrm{Tr}   \hat M_k \sin (\phi _k +\phi _0  )  =0.  
\end{eqnarray} 
Using Eqs. (\ref{form1}-\ref{to1}) we can rewrite Eq. (\ref{fc1}) as 
\begin{eqnarray}
F_c %& = & \frac{1}{2}\left(  1+ \sup_{\hat M_k}  \sum_k  \rho_k    \sqrt{   \gamma'^2 + (1- \gamma'^2 )  {\Delta _k }^2/ {\rho_k} ^2  }  \right) \nonumber\\
  &= & \frac{1}{2}\left(
  1+ \sup_{\hat M_k}  \sum_k   \frac{\textrm{Tr}\hat M_k  }{2}      \sqrt{ f( \phi_k)  }  \right)  \label{fc2}
\end{eqnarray}
where    
%\begin{widetext}
\begin{eqnarray}
 f( \phi )   
&\equiv & (1- {\gamma ' } ^2 ) (P+ G \cos \phi)^2 + \frac{{\gamma' }^2}{G ^2}  \times \nonumber \\ && \Big [ G + P \cos\phi -(1-P^2) \gamma \sqrt {1 -{\gamma }^2} \sin \phi  \Big ]^2 .  \nonumber \\   
\end{eqnarray}

In order to find an upperbound of $F_c$, let us consider 
a three-dimensional loop $\{ (x,y,z)=(\cos \phi, \sin \phi ,\sqrt{ f( \phi )}) | 0 \le \phi\ < 2 \pi\}$ and its tangent plane who has two points of tangency with $\phi = 0$ and $\phi = \pi$. %, $(1,0,\sqrt {f(0)} )$ and $(1,0,\sqrt {f(\pi)} )$. 
If we define another loop on the plane $\{ (x,y,z)=(\cos \phi, \sin \phi , { g( \phi )}) | 0 \le \phi\ < 2 \pi\}$ with% the component $g$ is given by 
\begin{eqnarray} g( \phi )   
&\equiv & K + \frac{1}{G K} \Big[ K^2   P \cos\phi \nonumber \\ && -{\gamma '}^2(1-P^2) \gamma \sqrt {1 -{\gamma }^2}  \sin \phi  \Big], \nonumber \\\end{eqnarray} and  \begin{eqnarray}
K&\equiv& \sqrt{  {\gamma'} ^2 +(1-{\gamma'}^2)G^2}, \end{eqnarray} then we can directly verify that the latter loop is always above the former one, that is, 
\begin{eqnarray} &&g( \phi ) ^2 -f( \phi )\nonumber\\ &=&  \frac{1}{K^2}( 1 -{\gamma '}^2 )(1-P^2) (1 -{\gamma }^2)\times \nonumber\\ &&  [(1 -{\gamma' }^2)G^2+ {\gamma '}^2 (1-(1-P^2)\gamma ^2)] \ge 0. \end{eqnarray} With this inequality and Eqs. (\ref{Pz}, \ref{Px}, \ref{fc2}), we obtain
\begin{eqnarray}
F_c   &\le  & \frac{1}{2}\left(
  1+ \sup_{\hat M_k}  \sum_k   \frac{\textrm{Tr}\hat M_k  }{2}   g( \phi_k)    \right) \nonumber\\ &=&  \frac{1}{2}\left(
  1+ \sup_{\hat M_k}  \sum_k   \frac{\textrm{Tr}\hat M_k }{2} K  \right)  =  \frac{1}{2}\left(
  1+  K  \right).  
\end{eqnarray} 
The upperbound is achievable by the POVM with two elements,  $\phi_k =0 $ and $\phi_k = \pi  $, which form spectral decomposition of $\hat  \Delta$.  Therefore, we obtain \begin{eqnarray}
F_c&= & \frac{1}{2}\left(
  1+  K     \right) %\nonumber \\  &= &
 =  \frac{1}{2}\left(
  1+   \sqrt{B   (2p_+ -1)^2  +  1- B }   \right)  \nonumber \\  \label{ec}\end{eqnarray} where we introduced the key parameter that represents the ``total non-orthogonalty'' of the state transformation % 
\begin{eqnarray}
B &\equiv&   (1- {\gamma ' }^2 ) {\gamma }^2   .  \label{b} \end{eqnarray}  
This quantity measures the non-orthogonality of the input states $\gamma$ with respect to the non-orthogonal axes $| \psi_\pm' \rangle$. %defined by the target states.
 When the target states are orthogonal $B$ reduces to $\gamma$ and $F_c$ corresponds to the success probability of MED for the two-state ensemble $\{p_\pm, |\psi_\pm\rangle \}$. 

It is worth noting that in the two-dimensional case the extreme EB map is \textit{classical-quantum} (CQ) map, that is, the measurement is orthogonal projection (see,  Th. 5 (D) of \cite{16}). Our approach here is in a sense to find the extreme point of EB maps. Hence, the same result will be obtained by restricting the optimization over CQ maps. Another approach for the optimization problem is found in a different context \cite{Bra07}. The optimization of $\bar F$ over CPTP maps is considered in \cite{Medo08}.

%\subsection{Criterion for Quantum domain processes with two input states}
\subsection{Criterion for quantum-domain transformation of two non-orthogonal states}
\begin{figure}[hbtp]%%%%%%%%%%%%%%%%%%%%%%%%%%%%%%%%%%%%%%%%%%%%%%%%%%%%%%%%%%
%{ABSA-%\includegraphics[width=8.0cm]{fig1.eps}
%\includegraphics[width=4.0cm]{fig1-conve.eps}
%\includegraphics[width=4.0cm]{fig1-witness.eps}
\includegraphics[width=8.6cm]{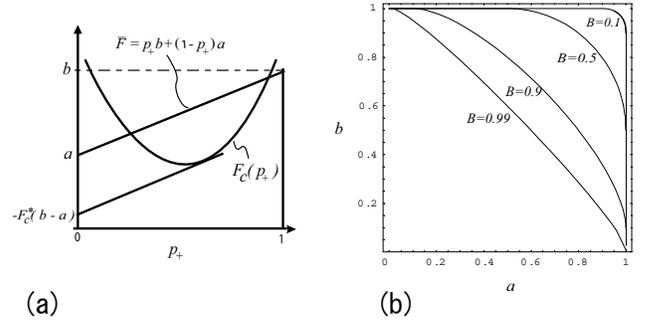}%{fig2-Leg1.eps }{ABSA-fig1.eps}{fig1-Leg.eps}
\caption{(a)The relation between the average fidelity $\bar F$ and the classical boundary $F_c$ as a function of the prior probability $p_+$. (b)The classical-quantum boundary for the fidelities $a$ and $b$ specified by Criterion (\ref{cr}) is shown for $B=0.1,0.5,0.9$, and $0.99$. \label{Legendre1}}
\end{figure}

Now we proceed to make the criterion for QD processes given the observed probabilities,  $b  = \langle \psi_+' |\hat \rho_+ |\psi_+' \rangle$ % =  \langle \psi_+' | E(|\psi_+  \rangle \langle \psi_+ |) |\psi_+' \rangle  $
 and $a  =  \langle \psi_-' |\hat \rho_- |\psi_-' \rangle$, %  = \langle \psi_-' | E(|\psi_- \rangle \langle \psi_- |) |\psi_-' \rangle  $.
%For simplicity we assume $ a \le  b  $.
where $ \hat \rho_\pm = E(| \psi_ \pm   \rangle \langle \psi_ \pm  | )$ is the output of the channel corresponds to the input $| \psi_\pm \rangle$.  
With the expression of the classical boundary fidelity $F_c$ of Eq. (\ref{ec}), the problem is the existence of $p_+$ that satisfies $\bar F = p_+ b + (1- p_+) a > F_c (p_+ )$. % whereas Sasaki and Fuchs considered the minimizatrion of $F_c (p_+ )$ with respect to prior distribution. %, \min_{p_+ + p_- =1} F_c$.
 Let us consider $\bar F $ and $\bar F_c $ as functions of $p_+$ (see FIG. \ref{Legendre1}a). Then we can see that $\bar F >F_c$ is satisfied if the segment that connects $(0,a) $ and $(1, b)$ is above the tangent line of $F_c$ whose slope is $b - a$. The condition is $a > - F_c^* (b-a)   $ where $F_c^* $ is the Legendre transform of $F_c $ defined by $F_c^*( \lambda ) \equiv \min_{1/2\le  p_+ \le 1} \{ \lambda p_+  -F_c(p_+)   \} $, noting that $F_c$  is convex.  
In this case, we can obtain $- F_c^* (b -a) = F_c(p_0) - (b-a) p_0$ where
$p_0 =\frac{1}{2}\left(1+ (b-a )\sqrt\frac{1- B}{B(B-(b-a)^2)} \right) $  is the solution of the equation $\frac{\partial}{ \partial  p_+}  F_c = b-a $.
From elementary calculation we obtain a simple QD condition in terms of the direct arithmetic mean $ \frac{a + b}{2} $, the slope $b-a$, and the overlaps $\gamma $ and $\gamma'$, those are in $B$ defined by Eq. (\ref{b}):  
\begin{eqnarray}
\frac{a+b}{2} &>& \frac{1}{2}\left( 1 + \sqrt\frac{(1-B)( B - (b-a )^2)}{ B}     \right)   \label{cr} .\end{eqnarray}
Typical behavior of the boundary with respect to $a$ and $b$ for various $B$ is shown in FIG. \ref{Legendre1}b. 
%\begin{figure}[hptb]%%%%%%%%%%%%%%%%%%%%%%%%%%%%%%%%%%%%%%%%%%%%%%%%%%%%%%%%%%
%\includegraphics[width=5cm]{domain2.eps}\caption{The boundary specified in the condition (\ref{cr}) is shown. \label{domain1}}\end{figure}

%\begin{figure}[hptb]%%%%%%%%%%%%%%%%%%%%%%%%%%%%%%%%%%%%%%%%%%%%%%%%%%%%%%%%%%
%\includegraphics[width=5cm]{noise-1.eps}\caption{The boundary specified in the condition (\ref{cr}) is shown. \label{bpsk}}\end{figure}

This criterion provides a relation between the change of ``purity'' and the change of non-orthogonality in order that the process maintains the inseparability (The fidelities give a lowerbound of the purity and operator norm of the output states such as $\textrm{Tr} \hat \rho _+^2 \ge a^2$ and $\| \hat\rho_+ \|_\infty \ge a$). 
Certainly, the criterion is satisfied if both of the input states preserves the purity, i.e., $a=b=1$.
Moreover, it is known that a qubit channel is EB channel if it transforms a pure two-qubit entangled states into a separable state \cite{Kinoshita}. % 
Hence, if our criterion is satisfied, we can fine a set of pure two-qubit entangled states whose inseparability survives after the local process. %

It might be valuable to consider %extend the criteria in 
the case where the input states are mixed states. %,  
Suppose that the input mixed states $\hat \sigma_\pm $ are prepared by subjecting a CPTP map $\mathcal L$ on a pair of pure states $|\phi_ \pm \rangle$. If the total process $E \circ \mathcal L $ is in QD, it is clear that $E$ is in QD. Then, we can use the criterion assuming the task $|\phi_ \pm \rangle \to |\psi_ \pm ' \rangle$. Since the condition that there exists a CPTP $\mathcal L$ satisfying $\hat \sigma_\pm  = \mathcal L (|\phi_ \pm \rangle \langle \phi_ \pm | )$ is $| \langle \phi_+ |\phi_ - \rangle | \le \textrm{Tr}\sqrt{ \hat \sigma_+ ^{1/2} \hat \sigma_- \hat \sigma_+^{1/2}   }$, we obtain the criterion in the case of mixed input by replacing $\gamma$ with the Uhlmann fidelity $ \textrm{Tr} \sqrt{ \hat \sigma_+ ^{1/2} \hat \sigma_- \hat \sigma_+^{1/2}   }$.

\section{applications}
\subsection{transmission of binary coherent states}
In quantum optical experiments, one of the most accessible state is the optical coherent state $| \alpha\rangle  = \hat D( \alpha ) |0 \rangle $ where $\hat D( \alpha ) =e^{\alpha \hat a^\dagger - \alpha ^* \hat a}$ is the displacement operator and $| 0\rangle$ is the vacuum state defined by $\hat a | 0\rangle =0 $. %  pair of non-orthogonal states  A commonly prepared One of the most 
In many situations, the ideal lossy channel is useful to describe the transmission of the coherent state as a first approximation. The ideal lossy channel with the transmission $\eta$ transforms the coherent state as  $ | \alpha \rangle \to | \sqrt \eta  \alpha \rangle $. This evolution preserves the purity and the ideal lossy channel is clearly in QD.
 A natural question is the maintenance of coherence in the presence of excess noise \cite{Rig06,Has08,namiki07,Gro03b,namiki04}. For the case of Gaussian-distributed input coherent states on the phase-space, one can find QD criteria where the noise is measured in terms of quadrature variance \cite{namiki07,Gro03b} or average fidelity \cite{namiki07}. %

  In order to apply our criterion for a lossy channel one may use the binary coherent state $|\psi_\pm \rangle = |\pm \alpha \rangle $ and choose $|\psi_\pm' \rangle = |\pm \sqrt \eta  \alpha \rangle $.  We can take $\alpha >0$ without loss of generality.
 The experimental data $a$ and $b$ are directly measured by the photon detection after appropriate phase-space displacement \cite{namiki07}. The threshold photon detector discriminates the  more-than-one-photon states from the vacuum state, and the measurement statistics give the probability of the projection to the vacuum state $\textrm{Tr} \hat \rho | 0\rangle \langle 0 | = \langle 0| \hat \rho | 0 \rangle$. 
 Hence, the photon detection after the displacement $\hat D (-\alpha)$ gives the projection probability to the coherent states $| \alpha \rangle$ so that $\textrm{Tr} \left[ \hat D (-\alpha)\hat \rho  \hat D^\dagger (-\alpha)|0 \rangle\langle 0|\right] = \textrm{Tr}  \left[ \hat \rho  \hat D(\alpha)|0 \rangle\langle 0| \hat D^\dagger (\alpha) \right] =\langle \alpha |  \hat \rho | \alpha \rangle$. Therefore, in this case, the verification scheme can be realized in a common quantum optical experimental technology of preparation of binary coherent states, displacement, and threshold photon detection.  
  
Since any minimum uncertain state is a pure state, the ``purity'' can be connected with the noise of the quadratures. Actually, it is possible to estimate a lower bound of $a$ and $b$ by homodyne measurements. % It .
We define the quadrature operators as $\hat x_1    \equiv  \frac{\hat a +\hat a ^\dagger }{2}$, $\hat x_2 \equiv \frac{\hat a - \hat a ^\dagger }{2i}$.  
Using Eq. (\ref{a1}) of Appendix A with $r=0 $, we have 
\begin{eqnarray}
\langle 0| \hat \rho | 0 \rangle \ge \frac{3}{2}  - \textrm{Tr} \left (\hat \rho (  \hat x_1 ^2 +\hat x_2 ^2  ) \right)   .
\end{eqnarray}
By substituting
%\begin{eqnarray}
$\hat \rho = \hat D^\dagger  (\pm \sqrt \eta \alpha ) \hat \rho_  \pm \hat D(\pm \sqrt \eta \alpha )$ into this inequality, 
%\end{eqnarray}
a lowerbound of $a$ and $b$ is given as 
\begin{eqnarray}
 \langle \pm \sqrt \eta \alpha | \hat \rho_\pm | \pm \sqrt \eta \alpha  \rangle &\ge& \frac{3}{2}  -\textrm{Tr} \left [\hat \rho_\pm  (  ( \hat x_1 \mp \sqrt\eta \alpha ) ^2 +\hat x_2 ^2  ) \right] .  \nonumber \\
%& = & \frac{3}{2}  -\textrm{Tr} \left [\hat \rho_\pm    \hat x_1  ^2  \right]  \pm 2 \sqrt\eta \alpha \textrm{Tr} \left [\hat \rho_\pm  \hat x_1   \right] -  \eta \alpha^2 -\textrm{Tr} \left [\hat \rho_\pm  \hat x_2 ^2  \right]     \nonumber \\ &\equiv  & \frac{3}{2}  -\langle \Delta^2 (\hat  x_1) + \Delta^2 ( \hat x_2) \rangle _\pm   \nonumber \\& &- \langle \hat x_1 \mp \sqrt \eta \alpha \rangle _\pm ^2- \langle \hat x_2 \rangle _\pm ^2  .
 \end{eqnarray}
% where  $\langle \  \bar \cdot \    \rangle _\pm  \equiv\textrm{Tr}(\hat \rho_\pm  \  \bar \cdot \   )$ and $\Delta^2 (\hat O)$ means the variance of $\hat O$. 
 The right hand side (rhs) of this inequality consists of the first and second moments of the quadratures, and is estimated by the homodyne measurements. 

There is a different QD criterion that directly concerns
quadrature noises \cite{Rig06,Has08}, which is formulated to verify the entanglement between an optical mode and a qubit, e.g., $|\alpha\rangle|0\rangle +|-\alpha\rangle | 1  \rangle $. This criterion requires the measurements of four observable $\hat x_1,\hat x_1^2, \hat x_2, \hat x_2^2$
for each of the two input states $|\pm \alpha\rangle$, and uses eight quantities. The derivation of the criterion is based on the negative partial transpose of the virtual entangled states. On the other hand, the present method is derived based on the optimization of M\&P schemes and requires measurements of one observable for each of the input states, i.e., we use only two quantities, $a$ and $b$.

\subsection{teleportation of single-photon polarization states}
An interesting application of the QD criterion is the proof of entanglement assistance for the quantum teleportation. In the Innsbruck first experiment of teleportation \cite{Bouw97}, the transmission of the two polarized single-photon states with the relative angle of $\pi /4 $ were considered. In this case we take $|\psi_\pm \rangle =|\psi_\pm ' \rangle  $, $ \gamma = \gamma ' = 1 / \sqrt 2$ and then $B = 1/4 $. The observed values of the fidelities are about $a = 0.82$ for $45 ^\circ$-polarized state and  $b =0.82$ for $90 ^\circ$-polarized state  \cite{Bouw99,Bouw-ed}. These fidelities are not high enough to satisfy Criterion (\ref{cr}). The requirement of high fidelities for the two-state scheme was already pointed out in \cite{Fuc03,Hend00}.

\subsection{storage of squeezed vacuum states}
There are broad approaches to show the quantum nature of the processes associated with degree of squeezing. Intuitively, the maintenance of squeezing suggests that the process will convey the signal with fine structures under the shot noise limit. %
Here, we are concerned with the squeezed light as a source of the non-orthogonal states and show how to apply our QD criterion to the experiments that provide the degrees of squeezing before-and-after the storage or transmission processes \cite{Honda07, Appel07, Yonezawa07}. The experiments are mainly initiated to realize a higher dimensional quantum information processing. The objective of the use of our criterion is to find a workable qubit subspace embedded in the higher dimensional space. 

If a squeezed vacuum state is generated, one can prepare a set of non-orthogonal states by applying phase rotations on the squeezed vacuum. Hence we consider a pair of thermal squeezed vacuum states connected with a phase rotation $\mathcal R_\theta $ as the two input, $\hat \sigma_+$ and $\hat \sigma_- \equiv  \mathcal R_\theta (\hat \sigma_+) $. We assume that the process $E$ is phase insensitive, i.e., $E \mathcal R_\theta = \mathcal R_\theta  E $ for any $\theta$ of the rotation angle. % 
Suppose that the transition of  $\hat \sigma_+$ due to the process transforms the covariance matrix of $\hat \sigma_+$ as \begin{eqnarray}
\mathcal{C}(\hat \sigma_+ )  &\equiv&   \left(  \begin{array}{cc }
X & 0   \\
0 &Y \\ 
\end{array}\right)\to\mathcal{C}(\hat \rho_+ )   \equiv    \left(  \begin{array}{cc }
X' & 0   \\
0 &Y' \\ 
\end{array}\right).  
\end{eqnarray}
Then the transition of the other state is given by %can be written as
\begin{eqnarray}
  \mathcal{C}(\hat \sigma_- )  & =&   \mathcal{C}(\mathcal{R}_\theta (\hat \sigma_+ )) \equiv R(-\theta )\mathcal{C}(\hat \sigma_+ )R(\theta ) \nonumber \\
     \to  \mathcal{C}(\hat \rho_- ) &=&  R(-\theta )\mathcal{C}(\hat \rho_+ )R(\theta ) \end{eqnarray} 
where %\begin{eqnarray}
$R(\theta ) \equiv   \left(  \begin{array}{cc }
\cos  \theta& - \sin  \theta  \\
 \sin  \theta&\cos   \theta\\ 
\end{array}\right)$. %\end{eqnarray}
We also assume the first moment of the quadratures for the output states vanish, $\textrm{Tr}[\hat \rho_\pm \hat a ] =0 $. 
 
A feasible choice of the target states is squeezed vacuums connected with the rotation. We write %use   $|\psi_+'\rangle = \hat S(r) |0\rangle $ and $|\psi_-'\rangle = \hat R(\theta )  |\psi_+ '\rangle $ represented in the terms of covariance matrices with 
the squeezing parameter $r$ and then the covariance matrices of the target is given by  \begin{eqnarray}
\mathcal{C}(|\psi_+'\rangle \langle \psi_+'| )  &\equiv&   \left(  \begin{array}{cc }
e^{2r} & 0   \\
0 &e^{-2r} \\ 
\end{array}\right), \\
\mathcal{C}(|\psi_-'\rangle \langle \psi_-'|  )  & \equiv&   R(-\theta) \mathcal{C}(|\psi_+'\rangle \langle \psi_+'| ) R(\theta).
\end{eqnarray} If the oputput $\hat \sigma_\pm$ is Gaussian state, the fidelity to the target state is given by  
\begin{eqnarray}
 \langle \psi_\pm'| \hat \rho_\pm  | \psi_\pm' \rangle  
&=&\frac{2}{\sqrt{e^{2r} Y' + e^{-2r}X'   +X' Y' +1 } } \nonumber \\ &\le&  \frac{2}{1+ \sqrt{X' Y'  }} \label{eq37} 
\end{eqnarray} where we use Eq. (\ref{fofg}) of Appendix B and the inequality comes from the relation of the geometric-and-arithmetic means. %the .
 Here $r$ can be selected to obtain the upperbound so that $e^{2r} = \sqrt{X'/Y'}$, and the fidelities are estimated by \begin{eqnarray}
a=b
 &=&  \frac{2}{1+ \sqrt{X' Y'  }}.\label{abxy}
\end{eqnarray} 
Using Eq. (\ref{fofg}) again, we have
\begin{widetext}\begin{eqnarray}
\gamma ^2 &=& \left(\textrm{Tr}\sqrt{\sqrt{\hat \sigma_+}   \hat \sigma_- \sqrt{\hat \sigma_+} }\right)^2 %&= &\frac{2}{\sqrt{(\det( \mathcal{C }(\sigma _+) +\mathcal C (\sigma_-) )+(\det \mathcal{C }(\sigma _+) -1) (\det \mathcal{C }(\sigma _-) -1)}-\sqrt{(\det \mathcal{C }(\sigma _+) -1) (\det \mathcal{C }(\sigma _-) -1)   }} \nonumber \\ 
 =   \frac{2}{\sqrt{X^2 Y^2+ \frac{1}{2}\left[ (X+Y)^2- (X-Y)^2 \cos (2\theta )  \right]+1} -XY +1}   \nonumber \\
{\gamma '} ^2 &=& |\langle \psi'_+| \psi'_- \rangle|^2= \frac{2}{\sqrt{2+ \frac{1}{2}\left[ (X'+Y')^2- (X'-Y')^2 \cos (2\theta )  \right] } } \label{eq39}.
\end{eqnarray}
 \end{widetext}

\begin{table}[htbp]
 \caption{The lhs of Criterion (\ref{cr}) is estimated from the degrees of squeezing (antisqueezing) for input states $X$ ($Y$) and for output states $X'$ ($Y'$) in experiments \cite{Honda07, Appel07, Yonezawa07}.  The last two columns are the minimized value of the rhs of Criterion (\ref{cr}) with respect to the rotation angle and the value of the angle that achieves the minimum, $\theta_{min}$. The criterion (lhs)$>$(rhs) is not satisfied.} 
 \begin{center}
  \begin{tabular}{|c|c|c|c|c|c|c|c|}
    \hline
  Ref.    &   $X$ (dB) & $Y$ (dB) & $X'$ (dB)   & $Y'$ (dB)   &  lhs & rhs   & $\theta_{min}$   \\
    \hline
 %  Honda \textit{et al.}
  \cite{Honda07} I  &  -2  &  6  & -0.07   &0.49    & 0.77   &  0.994  & 0   \\
    \hline
 %  Honda \textit{et al.}
  \cite{Honda07} II  &  -1.24  &  4.1  & -0.16   &0.90    & 0.84   &  0.989  & 0   \\
    \hline
%Appel \textit{et al.}
% \cite{Appel07}   &  -1.23  & 3.74   &  -0.26  &  1.77  & 0.86   %&  0.975  & 0  \\
%    \hline
%Appel \textit{et al.}
 \cite{Appel07}   &  -1.86  & 5.38   &  -0.21  &  1.32  & 0.80   &  0.983  & 0  \\
    \hline\cite{Yonezawa07}%Yonezawa \textit{et al.} 
  & -6.2   & 12.0   & -0.8   &12.4    & 0.68   & 0.800   & 0   \\
    \hline
  \end{tabular}
 \end{center}
\end{table}

Now we can directly evaluate the both sides of Criterion (\ref{cr}) for the experiments that investigate the degree of squeezing before-and-after the process. %
%\fbox{\parbox{ \linewidth}{ 
For the experiment demonstrated by Honda \textit{et al.} \cite{Honda07} (Method I), the degrees of squeezing and antisqueezing were $X=0.63$ (-2dB), $Y=3.98$ (6dB), $X'=0.98$ (-0.07dB), $Y'= 1.12$ (0.49dB), and $a=b=   \frac{2}{1+ \sqrt{X' Y'  }} =0. 77 $. %, we use Formula (\ref{cr}). 
With the help of Eq. (\ref{eq39}), the rhs of Ineq. (\ref{cr}) %, $ \frac{1}{2}(1+\sqrt {1-B} )$,
 is a function of $X$, $Y$, $X'$, $Y'$ and $\theta$, and is minimized to $ 0.994$ when $\theta = 0 $. The results of similar calculation for the experiments \cite{Honda07,Appel07,Yonezawa07}
 are summarized in Table I. Unfortunately, we have not found the result of the experiments where the process is supposed to have enough coherence to satisfy our criterion. %it is not shown that the experiments have enough coherence to satisfy our criterion.
 %}} 

Note that the output-to-target fidelity of Eq. (\ref{eq37}) is for the Gaussian states. In realistic, it is not always reasonable to assume that the states are Gaussian. In such case, we can use the lowerbounds estimated from the quadrature measurement given in Appendix A. %it i  
If we choose $ e^{2 r }= \sqrt{\frac{ \textrm{Tr}   (\hat \rho    \hat x_1  ^2 )}{\textrm{Tr}   (\hat \rho    \hat x_2   ^2 )  }}  $ in Eq. (\ref{a1}), the projection probability is lowerbounded by the observed quadrature noises as  
%\begin{eqnarray}
$\langle \psi_\pm '|\hat \rho_\pm | \psi_\pm ' \rangle% = \langle 0|\hat S^\dagger (r)  \hat \rho \hat S (r) | 0 \rangle 
\ge \frac{3}{2}  - 2 \sqrt{ \textrm{Tr} (\hat \rho_\pm  \hat x_1 ^2     ) \textrm{Tr} ( \hat \rho_\pm  \hat x_2 ^2 ) }$. % \nonumber \\\end{eqnarray}
Hence, we can use \begin{eqnarray}a=b \ge \frac{1}{2}(3 - \sqrt{X'Y'})\end{eqnarray} %$a=b \ge \frac{1}{2}(3 - \sqrt{X'Y'})$ 
instead of Eq. (\ref{abxy}) provided $\textrm{Tr} (\hat \rho_\pm  \hat x_1      )= \textrm{Tr} (\hat \rho_\pm  \hat x_2     ) =0 $.

\section{Conclusion}
We have considered the average fidelity of the transformation task between two pairs of non-orthogonal pure states for a given quantum channel and derived a QD criterion The criterion takes simple form with a few experimental parameters and provides a relation between the change of ``purity'' and the change of non-orthogonality in order that the channel maintains the inseparability. The criterion can be applied for the case of mixed input states by using % 
the Uhlmann fidelity between the mixed inputs.
We made a few examples of applications for quantum optical experiments. In particular, we showed how to apply our criterion
for the experiments of storage or transmission of squeezed states. % 
 While the criterion provides a concrete foundation on the transfer of an unknown quantum state in relation with the non-orthogonality, it is likely that surpassing the classical boundary achievable by the classical M\&P device requires higher fidelities and lower-noise experiments than the achievement of the present experiments.
It will be valuable both in fundamentally and technologically to establish quantum channels that attain such a high-standard benchmark.

\acknowledgements
The author thanks M. Koashi and N. Imoto for helpful discussions. 
R.N. is supported by JSPS Research Fellowships for Young Scientists.

%This work was supported by a MEXT Grant-in-Aid for Young Scientists (B) 17740265.

\appendix
\section{Measurement of fidelity to squeezed state}
The fidelity to a coherent state can be given by the photon detection followed by displacement as described in the Sec. IIIA. Similarly the fidelity to a squeezed state can be given by the probability of photon detection after certain displacement and squeezing. While the former can be realized within standard technique of linear optics, the latter requires squeezing operation. % 
In this appendix we provide a method for estimating the fidelity to a squeezed state with linear optics and homodyne detection. 

Let us write the photon number operator $\hat n = \hat x_1^2 + \hat x_2^2-\frac{1}{2}$ and squeezing operator $\hat S(r)$ with degree of squeezing $r$ whose action to the quadrature operator is given by $\hat S^\dagger (r)(\hat x_1+ i\hat x_2 )\hat S(r)=\hat x_1  e^{ r} + i \hat x_2  e ^{-r }$. We define a squeezed photon number operator by  %$\hat n =x^2 + p^2$ 
$\hat n_{ S}(r)  \equiv  \hat S^\dagger (r) \hat n \hat S(r)$. % = x^2 e^{2 r} + p^2 e ^{-2r }$. 
Using the spectra decomposition of $\hat n = \sum_{n=0}^{\infty} n |n \rangle \langle n|   $, % we have $\hat n_s = \sum_{n=0}^{\infty}n \hat S^\dagger   |n \rangle \langle n|\hat S    $ and 
we can see that
$\textrm{Tr} (\hat \rho \hat n_S )= \sum_{n=1 }^{\infty}n \langle n|\hat S^\dagger \hat \rho \hat S |n\rangle \ge
\sum_{n=1}^{\infty} \langle n|\hat S^\dagger \hat \rho \hat S  |n\rangle = 1-\langle 0|\hat S^\dagger  \hat \rho \hat S | 0 \rangle
$
for any normalized state $\hat \rho$. The inequality comes from $n\ge 1$.  Hence, we have
\begin{eqnarray}
\langle 0|\hat S^\dagger (r)  \hat \rho \hat S (r) | 0 \rangle &\ge& 1  - \textrm{Tr} \left (\hat \rho \hat n_S(r ) \right) \nonumber\\ %(  \hat x ^2 e^{2 r } +\hat p ^2 e^{-2r} )
&=& \frac{3}{2}  - \textrm{Tr} \left (\hat \rho    \hat x_1 ^2  \right) e^{2 r }- \textrm{Tr} \left (\hat \rho x_2 ^2   \right)e^{-2r}.\nonumber \\ \label{a1}\end{eqnarray} % 
This provides a lower bound of the fidelity to the squeezed vacuum state  $\hat S(r)| 0 \rangle $ from the quadrature moments determined by homodyne measurements, $\textrm{Tr} \left (\hat \rho \hat x_1 ^2   \right)$ and  $\textrm{Tr} \left (\hat \rho \hat x_2 ^2   \right)$. By taking proper displacement on $\hat \rho$ beforehand, we obtain an estimation of the fidelity to any pure quadrature-squeezed state. %s that $\textrm{Tr} \left (\hat \rho x  \right)=\textrm{Tr} \left (\hat \rho p \right)=0$

As a function of $r$, the rhs of Eq. (\ref{a1}) is maximized when $ e^{2 r }= \sqrt{\frac{ \textrm{Tr}   (\hat \rho    \hat x_1  ^2 )}{\textrm{Tr}   (\hat \rho    \hat x_2   ^2 )  }}  $. This provides the  choice of the target state in the last part of Sec. III C.%where% .

\section{Covariance matrix and Fidelity between Gaussian states}
The covariance matrix for density operator $\hat\rho$ is defined by %$[\mathcal C(\hat\rho) ]_{1,1} =  4\left\{  \textrm{Tr}[\hat \rho  \hat x^2  ]  -[ \textrm{Tr}(\hat \rho  \hat x]^2 \right\} $--  
$ [\mathcal C(\hat\rho) ]_{i,j} =  4\left\{ \frac{\textrm{Tr}[\hat \rho  ( \hat x_j \hat x_k + \hat x_k \hat x_j  )] }{2} -\textrm{Tr}(\hat \rho  \hat x_j ) \textrm{Tr}(\hat \rho \hat x_k  )   \right\} $ with $i =\{1,2\}$.
The Uhlmann fidelity between Gaussian states $\hat \rho_1 $ and $\hat \rho_2$ 
 is given by \cite{Scu98},
\begin{eqnarray}
%\nonumber F (\hat \rho _1, \hat \rho _2)&\equiv& 
\nonumber &&\textrm{Tr}\sqrt{\sqrt{ \hat \rho _1} \hat  \rho _2 \sqrt{\hat  \rho _1 }} \\&=& 
 \nonumber  
 \sqrt{\frac{2}{\sqrt{\Delta + \delta }-\sqrt{\delta} } } \exp\left[  - \Lambda ^{ T}(\mathcal C (\hat \rho_1) + \mathcal C (\hat \rho_2) )^{-1} \Lambda   \right] %\nonumber,
 \\   \label{fofg}
\end{eqnarray}
where \begin{eqnarray}
% \Gamma_i &\equiv& \mathcal C (\hat \rho_i) \\ 
\Delta &\equiv& \det (\mathcal C (\hat \rho_1)  + \mathcal C (\hat \rho_2) ) ,\\
\delta & \equiv& (\det \mathcal C (\hat \rho_1)  -1  ) ( \det \mathcal C (\hat \rho_2)  -1 ),\\
%\\
\Lambda &\equiv&   
\left(\begin{array}{c}\textrm{Tr} ( \hat \rho _1   \hat x_1 )     -\textrm{Tr} (   \hat \rho _2  \hat x_1  )\\\textrm{Tr} ( \hat \rho _1  \hat x_2 )  - \textrm{Tr} (   \hat \rho _2  \hat x_2  )\end{array}\right). % \\  \Gamma_i  &\equiv& \left(\begin{array}{cc} (\Gamma_ i)_{11} &  (\Gamma_i)_{12} \\ (\Gamma_i)_{21}   & (\Gamma_i)_{22} \end{array}\right), \nonumber \\
\end{eqnarray} 
%\begin{eqnarray}


\begin{thebibliography}{}
\bibitem{qdc}If a quantum channel $E$ is in quantum domain, $\openone_A\otimes  E_B ( \hat \Psi_{AB})$ is an entangled state where $\hat \Psi$ represents the maximally entangled state with $\textrm{Tr}_A  \hat \Psi_{AB}  = \openone_B $ and $ \openone_B$ is the support of $E$.
\end{thebibliography}
\end{document}